# Generation of ultrafast electron bunch trains via trapping into multiple periods of plasma wakefields

Grigory Golovin[1], Vojtěch Horný[2,3], Wenchao Yan[1,4], Colton Fruhling[1], Daniel Haden[1], Junzhi Wang[1], Sudeep Banerjee[1], and Donald Umstadter[1]

[1] Department of Physics and Astronomy, University of Nebraska-Lincoln, Lincoln, Nebraska 68588, USA

[2] Chalmers University of Technology, SE-412 96 Gothenburg, Sweden

[3] Institute of Plasma Physics, Czech Academy of Sciences, Za Slovankou 1782/3, 182 00 Praha 8, Czech Republic

[4] Institute of Physics ASCR, v.v.i. (FZU), ELI BEAMLINES, Na Slovance 1999/2, 182 21 Prague 8, Czech Republic

## Abstract

We demonstrate a novel approach to the generation of femtosecond electron bunch trains via laser-driven wakefield acceleration. We use two independent high-intensity laser pulses, a drive, and injector, each creating their own plasma wakes. The interaction of the laser pulses and their wakes results in a periodic injection of free electrons in the drive plasma wake via several mechanisms, including ponderomotive drift, wake-wake interference, and pre-acceleration of electrons directly by strong laser fields. Electron trains were generated with up to 4 quasi-monoenergetic bunches, each separated in time by a plasma period. The time profile of the generated trains is deduced from an analysis of beam loading and confirmed using 2D Particle-in-Cell simulations.

## Introduction

The acceleration of electrons by laser wakefields (LWFA) is a rapidly maturing field. It is attractive because of its ultra-high acceleration gradient (two to three orders of magnitude higher than that of conventional radio-frequency accelerators), and its ability to generate electron bunches with ultra-short (femtosecond)[1] pulse duration, small energy spread, and low beam emittance[2]. However, recent advances in the area of controlled injection[3–7] make the generation of not only single ultrashort electron bunches, but also their doublets[4] or even trains.

Trains of ultrashort electron bunches have many applications such as pump-probe studies[8,9], super-radiant radiation[10,11], and resonant excitation of beam-driven wakefields[12]. They can be used to reduce space-charge effects that lead to deterioration of bunch divergence, transverse size, and length after it leaves the accelerator and travels in free space with no confining forces[2]. A bunch train, compared to a single bunch, can carry more charge before space-charge effects start to play a role since the charge is distributed over several separate femtosecond-long bunches.

Various ways to generate trains of ultrashort electron bunches have been proposed for conventional radio-frequency accelerators, including masking[13], transverse-to-longitudinal phase-space exchange[14], an inverse-free-electron laser with a chicane[15], irradiation of a photocathode with a comb-like laser pulse[16], and laser beatwave[11]. Laser-driven wakefields allow for new unique ways to generate bunch trains. Several possible schemes have been proposed recently and validated through simulations. A bunch train can be generated via periodic self-injection triggered by waist oscillation of a laser pulse propagating in plasma[17]. Two copropagating laser pulses of different frequencies can create a beatwave, resulting in periodic ionization-assisted injection[18], or add incoherently, yielding a periodic self-injection due to the oscillation of the size of the plasma bucket[19]. Similar oscillations can be induced by propagating a laser pulse with a significant amount of negative chirp[20,21]. Multi-bunch injection can occur on a density down-ramp[22]. Electron bunch trains were experimentally generated via optical beating injection[23] and near-threshold self-injection[23–25]; their time signatures were measured via coherent transition radiation (CTR). While all the aforementioned works utilized single injection techniques, recent experiments demonstrated that it is possible to combine techniques to generate doublets of independently controllable electron bunches[4,8]. These controllable bunches can then be converted into doublets of independently controllable x-ray pulses via inverse-Compton scattering[8].

In this work, we present a novel all-optical approach to the generation of ultrashort electron bunch trains in a laser-driven wakefield via the recently-demonstrated injection mechanisms of ponderomotive drift, wake-wake interference, and laser field pre-ionization[26]. We launch two high-intensity laser pulses, drive and injector pulses, and overlap them in plasma at a steep angle. Both pulses are strong enough to drive their own wakes. The interaction between the pulses and their wakes results in the injection of the free plasma electrons into the drive wake. The accelerated electron beams have quasi-monoenergetic multi-peaked spectra. Our 2D Particle-In-Cell (PIC) simulations show that multiple buckets of the drive wake get loaded with electrons in each bucket forming individual bunches, each separated by a plasma period. The multi-bucket character of injection is deduced based on an analysis of beam loading in the experimental data.

## Experimental setup

The experiments were performed at the University of Nebraska – Lincoln Extreme Light Laboratory, with the Diocles laser system, using an experimental setup similar to that discussed in Ref.[26]. The laser beam was split into two beams, driver and injector, with each beam compressed independently. The drive beam (1.2 J, 36 fs) was focused by an f/14 parabolic mirror to a 20±1 μm (FWHM) focal spot ($a_0 = 1.4 \pm 0.1$ in vacuum). The injector beam (0.9 J, 34 fs) was focused by a f/2 parabolic mirror to a 2.8±0.1 μm (FWHM) focal spot ($a_1 = 9.0 \pm 0.5$ in vacuum). The beams were polarized in the horizontal plane (in which they propagated) and intersected at a 155° angle inside a 2-mm gas jet. The collision angle was dictated by the restriction of the experimental setup. The electron beam energy spectra were measured using a magnetic spectrometer with a 0.7-T, 15-cm-long magnet and a fluorescent screen (fast Lanex) 23-cm downstream. To resolve multiple peaks in electron spectra and deconvolve them from beam divergence, we used an iterative algorithm[27].

# Experimental results

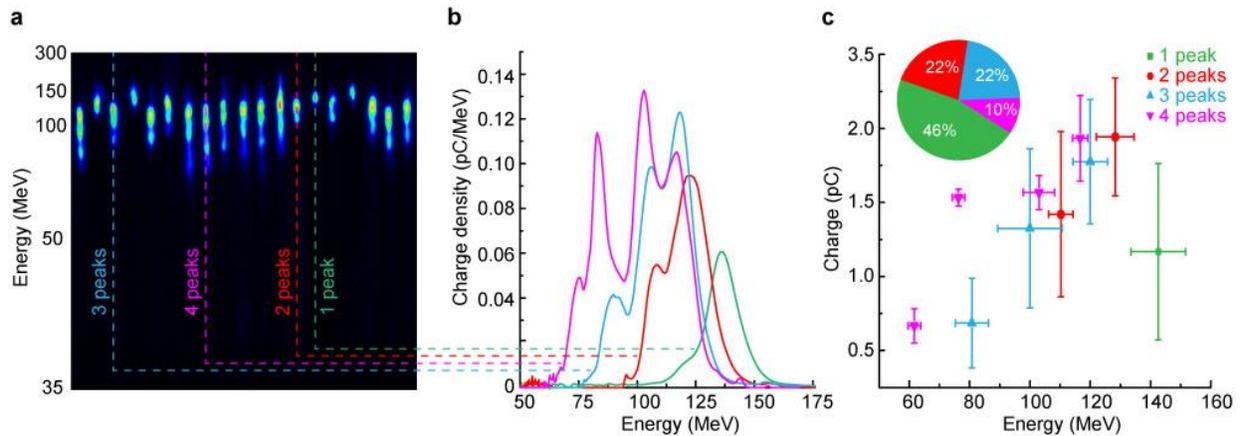

*Figure 1. a) Magnetically dispersed Lanex images of a series of electron beams measured in 19 consecutive laser shots. b) Representative spectra of the beams with 1, 2, 3, and 4 energy peaks from the series. c) Charge and central energy of individual spectral peaks of single- and multi-peaked electron beams for a larger, but non-consecutive series of 50 laser shots. The inset shows proportions of shots with electron beams with a different number of spectral peaks. For all shots shown, the plasma density is $0.76 \times 10^{19}$ cm$^{-3}$, drive and injector laser pulses normalized vector potentials are $a_0 = 1.4$ and $a_0' = 9$ (in vacuum).*

We first found a set of experimental parameters that resulted in stable injection via the mechanisms of ponderomotive push and wake-wake interference. The plasma density was scanned starting at $0.65 \times 10^{19}$ cm$^{-3}$, below which no injection of any kind was observed, and up to self-injection threshold at $1.3 \times 10^{19}$ cm$^{-3}$. In this range of plasma densities and with the drive pulse alone, we observed electron beams with negligibly low charge of 80±40 fC which were injected due to marginal wave-breaking over a short distance. With both drive and injector pulses, we observed electron beams with 1-2 orders of magnitude higher charge. Multi-peaked electron beams were present at all densities in this range, but the optimum was found at $0.76 \times 10^{19}$ cm$^{-3}$. We scanned the delay between the drive and injector laser pulses at this density in the range of -66 to +66 fs (with a plasma period of 40 fs) and observed beams with multiple energy peaks at every point of this scan. Figure 1a demonstrates a typical series of consecutive shots at these optimal conditions. Representative deconvolved spectra are shown in Figure 1b. One can see that the beams have a distinct multi-peaked spectral structure with 1-4 quasi-monoenergetic peaks at different energies. Statistics on energy and charge of the spectral peaks are shown in Figure 1c. Multi-peaked beams constituted more than half of all beams measured at the optimal experimental conditions.

We attribute fluctuation in the number of spectral peaks to two factors. The first factor is fluctuation in the intensity of the drive and injector pulses. While the measured level of these fluctuations in vacuum for both pulses is 5% (which includes fluctuations in laser power and focal spot size), it is most likely higher in the experiment due to non-linear character of laser pulse propagation in plasma. As it will be shown in the following simulation section, higher intensity of the injector pulse (as well as higher intensity of the drive one even though it is not shown in the paper) leads to injection of a single electron bunch, while lower intensity of the injector pulse (and lower intensity of the drive one) – to injection of multiple electron bunches. The second factor is pointing instability of the drive (standard deviation of 11 μm with 20-μm focal spot size) and injector (standard deviation of 2.0 μm with 2.8-μm focal spot size) laser pulses, which translates into different quality of their overlap. When the pulses overlap perfectly in the plane in which they propagate, the effective strength of the injector pulse, as seen by the drive pulse and its wake, is the highest. Again, it corresponds to a single injected electron bunch and hence, a single spectral peak in the measured electron spectra. When the drive and injector pulses do not overlap perfectly, the effective strength of the injector pulse is lower, and more electron bunches are injected in the drive wake, which leads to multi-peaked electron spectra.

We now address the question of the time profile of the electron beams with multi-peaked spectra. There are two possible scenarios leading to two different time profiles. First, a periodic injection can occur in the first period, or bucket, of the drive wake[17,19,20,22,24,18]. In this case, each injected bunch will be accelerated for a different amount of time and will, therefore, gain different energy. The bunches will be located inside a single bucket, and the total length of the bunch train will be less than a plasma wavelength. Second, the injection can occur in multiple periods, or buckets, of the drive wake[23,25]. In this case, each injected bunch will experience different accelerating gradient and gain different energy. The bunches will be separated by a plasma wavelength.

The most reliable way to find which injection scenario happened in the experiment would be to directly measure time profile of the electron beams, for example via Coherent Transition Radiation (CTR) technique[23,25]. However, one can get an insight into the electron beam time profile indirectly via the analysis of beam loading in the experimental data. As we will show, such analysis supports the multi-bucket injection hypothesis.

Beam loading occurs when the electric field of the injected charge is comparable with the field of the wake itself, and the accelerating gradient of the wake becomes modified[28,29]. When the injected charge grows, the modification gets stronger, the accelerating gradient drops, and the final energy of accelerated electrons becomes lower. Figure 2a shows the dependence between the energy of the accelerated electron beams and their charge for single-peaked electron beams. As expected, the energy drops when the charge grows. To statistically quantify the correlation between charge and energy, we used linear regression, as shown in Figure 2a, with a red line. For single-peaked electron beams, the p-value of the regression is $6.5 \times 10^{-6}$; since it is significantly less than 0.05 (commonly accepted level of significance), it means that the anti-correlation is statistically significant. We now move to multi-peaked beams. Figure 2b shows a similar dependence between the energy of the $1^{st}$ (highest energy) spectral peak and its charge; the p-value of the linear regression is $7.1 \times 10^{-6} \ll 0.05$, which again means that the anti-correlation is statistically significant. When we analyze how the same energy depends on the total charge, in all other spectral peaks, we see that these quantities are not related. The slope of the linear regression is zero within the error bars, the p-value is $0.93 \gg 0.05$ (the correlation is not statistically significant). This can be understood if the spectral peaks come from electron bunches situated in different periods of the drive wake. In this case, only the charge injected in the first period will affect the accelerating gradient of this period, and all the charges in later periods will not. The same trends can be seen on the dependences between the energy of the $2^{nd}$ spectral peak versus total charge in the peaks ##1 and 2 (Figure 2d, where the p-value is $5.2 \times 10^{-5}$) and peaks ##3 and 4 (Figure 2e, where the p-value is 0.7). The accelerating gradient of the $2^{nd}$ period is affected by the charges injected only in this period and the preceding periods (periods ##1 and 2), but not in the later periods (##3 and 4). If the other hypothesis were correct and all the charge is situated within the $1^{st}$ bucket, we would expect to see statistically significant negative trends in all these dependencies.

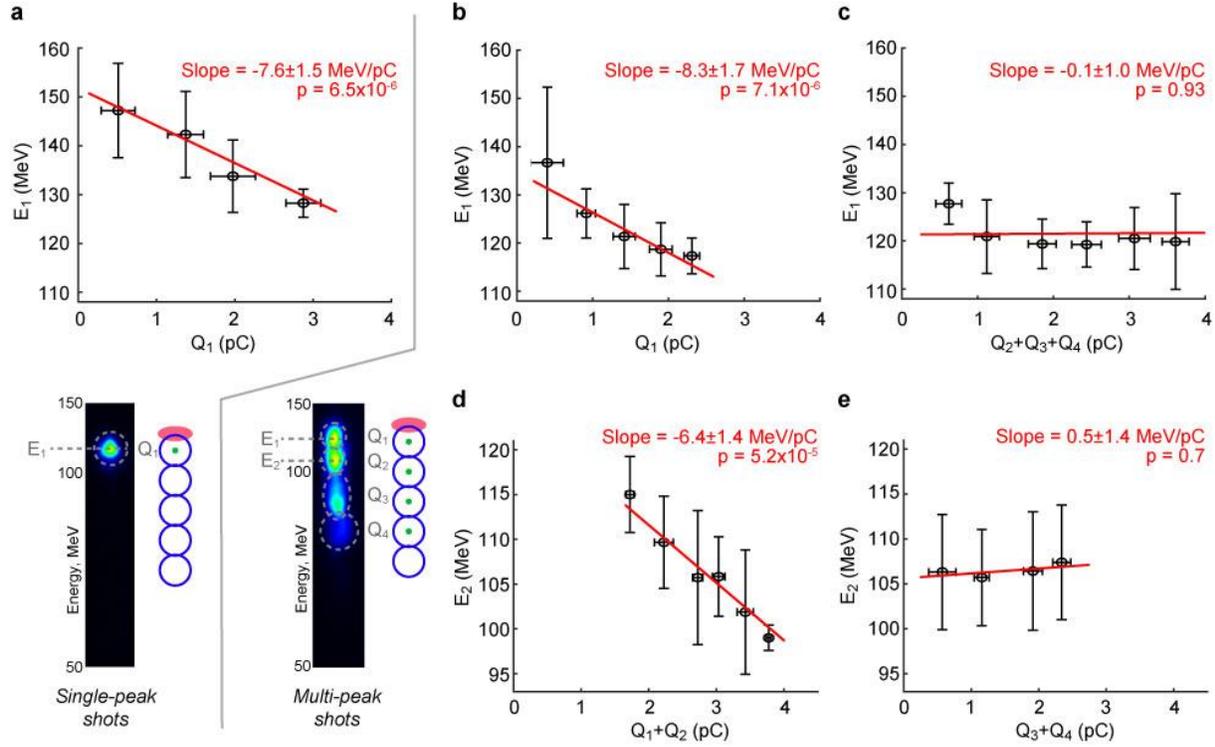

*Figure 2. Effects of beam loading on the energy of the accelerated electron beams. a) Shots with single-peak electron spectra. The energy of the spectral peak as a function of its charge. b-e) Shots with multi-peak electron spectra. Energy of the 1st peak as a function of its charge (b) and the sum of charges of all other peaks (c); energy of the 2nd peak as a function of the sum of charges of the 1st and 2nd peaks (d) and the sum of charges of the 3rd and 4th peaks (e). Data are binned, vertical, and horizontal error bars show standard deviations within the bins. Red lines show linear regressions. The insets at the bottom left show examples of dispersed Lanex images of single- and multi-peaked beams. The plasma density is 0.76x10^19 cm^-3, drive and injector laser pulses normalized vector potentials are $a_0$=1.4 and $a'_0$=9 (in vacuum). (a) contains 52 shots, (b), (c), and (d) – 51 shots, (e) – 35 shots.*

## Simulations

The process of injection and electron acceleration was also investigated with PIC simulations using a 2D version of the EPOCH code[30]. We chose two simulations with the following parameters to demonstrate the processes leading to the multibucket injection. The plasma wave drive pulse enters the simulation box from its left boundary. It is approximated as a Gaussian beam in both temporal and spatial domain, its laser strength parameter is $a_0 = 2$, duration $\tau_0 = 24$ fs, and waist size $\omega_0 = 14$ μm. The injection beam enters from the right boundary, and its parameters are laser strength $a'_0$ ranging from 2 to 9, duration $\tau'_0 = 29$ fs, and waist size $\omega'_0 = 4.8$ μm. These parameters represent a certain approximation which neglects the plasma effects on the beam temporal and spatial profile. Both beams are polarized in the horizontal xy plane and intersected at a 155° angle at the y-axis, 90 μm from the left boundary. Both pulses meet in the same spatial location. The plasma is represented as an electron gas with density 5x10^18 cm^-3 and a10-μm linear density ramps added normal to both beams propagation directions from practical purposes, which does not influence the physics of interaction. Considering the atypical configuration of the injection scheme, a rather large simulation box of $230 \times 145$ μm² is used. In the transverse direction, the box range is [-100 μm, 45 μm], and the drive pulse propagates along the y=0. The grid resolution is 36 and 22 cells per wavelength in the drive pulse propagation direction x and the transverse direction, respectively. Only the electron macroparticles are simulated as ions are assumed as the static background. Initially, four macroparticles are placed in every cell. A sixth-order solver on Yee's grid is used to integrate the Maxwell's equations. In general, 2D simulations tend to underestimate laser pulse self-focusing during its propagation through plasma[31]. While it is commonly accepted that the structure of the forces acting on the electrons inside the wakefield accelerator remains the same in both 2D and 3D geometries at least in the first bucket[32], possible inaccuracies of the wakefield strength due to 2D geometry might result in a slightly different final energies of the electron bunches. However, our intention is not to precisely replicate the experimental findings with simulations, but to capture the physics of injection.

Figure 3 and Figure 4 show the injection and acceleration phases in the snapshots of the electron density for $a'_0 = 2$ and $a'_0 = 9$, respectively. Only the central parts of the large simulation window are shown. Points of different colors illustrate positions of macroparticles which are eventually trapped into the first to the fifth period of the plasma wave, respectively. Before the collision of the laser pulses, the wakefield structure driven by the drive pulse contains a stable sequence of several periods. Then it is disturbed due to the interaction with the injection pulse and its wake. The extent of such disturbance differs significantly between the two cases. In the case with low-intensity injection pulse, the wakefield structure is still well defined after the collision, as it is shown in times of 0.8 ps and 0.9 ps in Figure 3. On the contrary, for the case with $a'_0 = 9$, the injection pulse disrupts the wake wave behind the drive pulse, and it takes about 1 ps before it restores its shape.

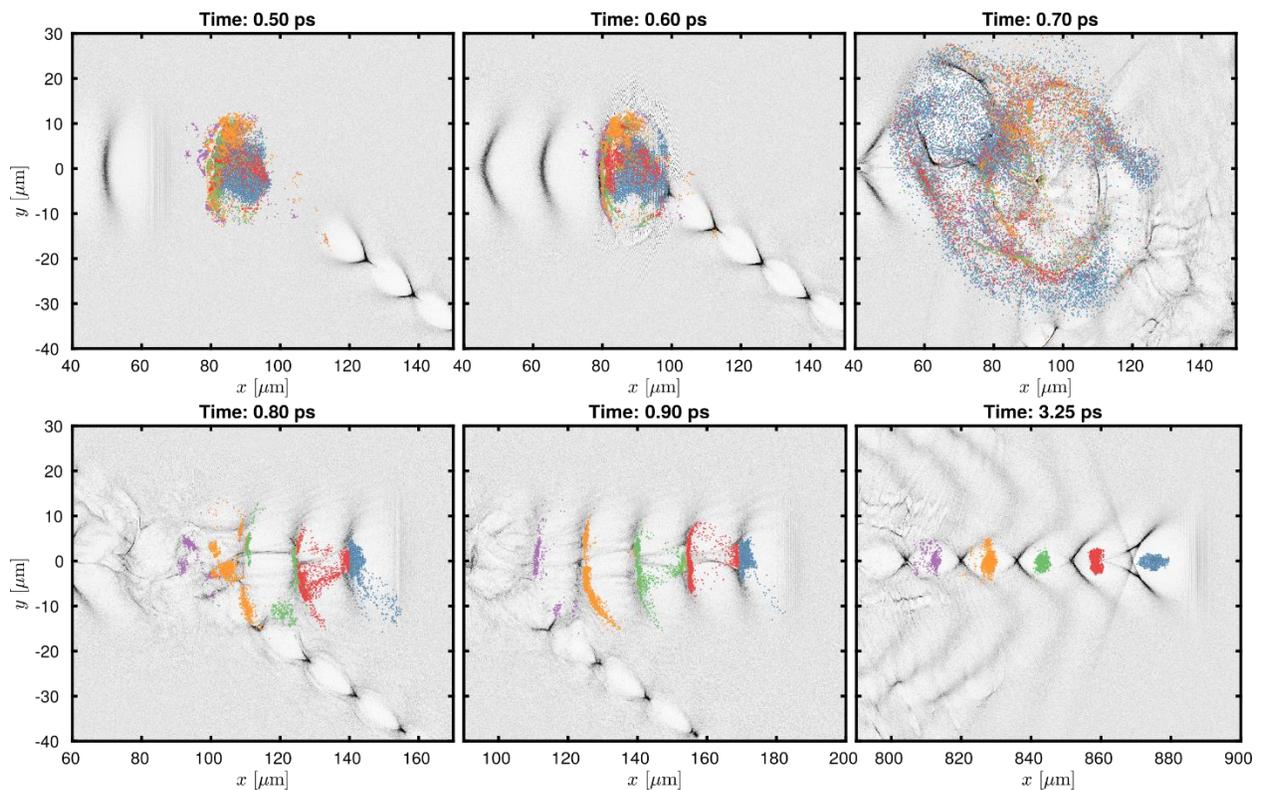

Figure 3. Density snapshots from PIC-simulations showing the evolution of the injection process, $a'_0 = 2$. Grey color scale represents the electron density in $[0, 5n_e]$ range, color points show the positions of electrons captured in the first five buckets.

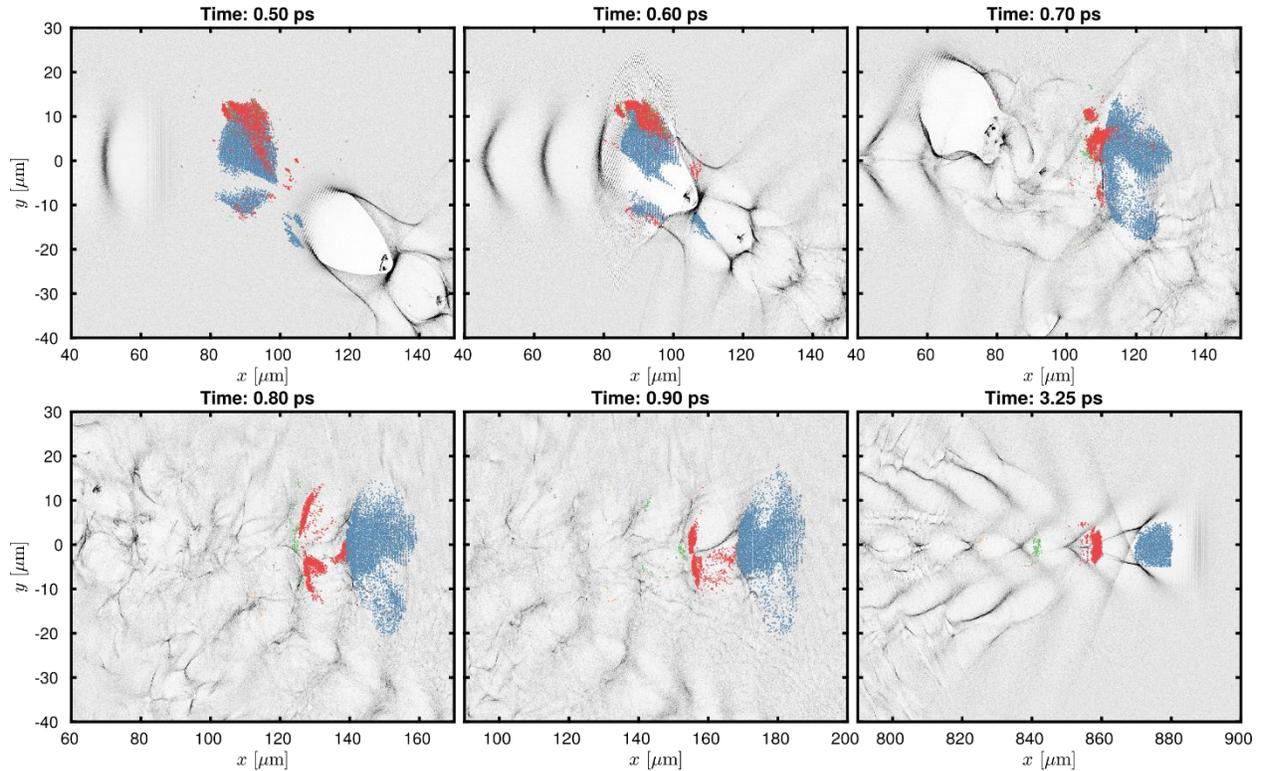

*Figure 4. Density snapshots from PIC-simulations showing the evolution of the injection process, $a'_0 = 9$. Grey color scale represents the electron density in $[0, 5n_e]$ range, color points show the positions of electrons captured in the first five buckets.*

The other difference between the low- and high-intensity injector cases is in the shape of the wake behind the injector pulse. In the low-intensity case (Figure 3), the wakefield is comprised of several well-defined plasma wave periods. As it can be seen from the electron density snapshots (times of 0.8 ps and 0.9 ps), the collision between the two wakes enhances the injection into the following buckets of the wakefield behind the drive pulse. On the other hand, the strong injection pulse wakefield comprises of well defined first two periods and scattered following periods (see Figure 4). Therefore, and also due to the great disturbance of the wakefield behind the driver, only a small amount of charge is trapped into the third and following buckets of the drive wake.

Figure 5 shows electron spectra for two different intensities of the injector pulse. The saw-like shape is visible in both cases. The colors of the curves correspond to the colors of the electrons trapped in different buckets in Figure 3 and Figure 4. For $a'_0 = 2$, four peaks in the energy spectrum correspond to electron bunches trapped in the first four buckets. The energy of the fifth bunch is about 15 MeV thus, the spectrum of that bunch is hidden within the background. The energy of the single bunches decreases with their distance to the driver pulse because the accelerating gradient decreases as well. This case is similar to the situation when multi-peaked electron beams were detected in the experiment. For $a'_0 = 9$, there are two distinct peaks in the spectrum coming from the first two buckets. Given that the majority of the charge comes from the first bucket, and the low-energy peak corresponding to the second bucket has a very low energy which is under the threshold of experimental detection, this high-$a'_0$ case is similar to the situation when single-peaked electron beams were detected in the experiment.

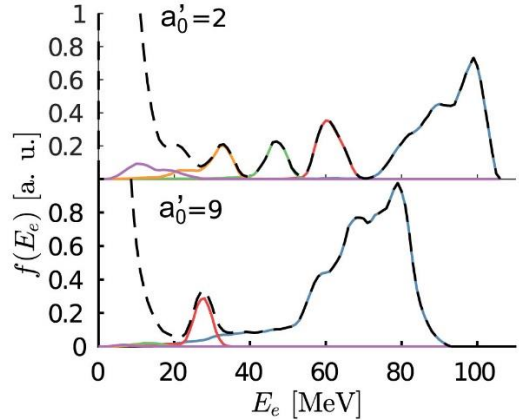

*Figure 5. Spectra of the accelerated electrons from PIC-simulations at the simulation time of 3.25 ps (2.65 ps after the collision of the laser pulses when the acceleration process has already stabilized). The color lines correspond to the groups of electrons trapped in the first five buckets which are shown in Figure 3 and Figure 4. Black dashed line shows the total energy spectrum comprising of all the macroparticles from the simulation box. The energy spectra are normalized according to the highest-energy peak of the bottom panel ($a_0' = 9$).*

We can identify various mechanisms leading to the injection of electrons. In general, it can be claimed that the electrons that would flow around the bubble without getting trapped in the case without injection pulse present are dephased from their fluid trajectories. Thus, some of them are captured. The processes resulting in that injection can be categorized as crossing beatwave injection[33], injection by laser field pre-acceleration[33], injection by the longitudinal momentum kick[34,35] or stochastic heating[36] in the standing wave, injection by a ponderomotive drift[37], wake-wake interference[26], and interaction of the injection pulse with the wake behind the driver. Also, even though most of the trapped electrons originate from the interaction region, a small fraction of them are shifted to the wake behind the drive pulse by the injection pulse or by its wake.

We found that the dominant mechanism responsible for electron injection into the first bucket in the case of $a_0'$=2 was the crossing beatwave injection, similarly to Ref. [33]. Electrons accelerated in the second bucket are often initially trapped in the first one, and later gradually move back. Electron injection in the third and following buckets is mainly due to the interaction of the injector pulse with the drive wake and also wake-wake interaction. In the case of $a_0'$=9 crossing beatwave injection is suppressed, and a mechanism similar to injection by laser field pre-acceleration of Ref. [33] dominates for the injection into the first bucket. However, in this case the pre-acceleration is also enhanced by the electrostatic fields of the bubble behind the injection pulse.

## Conclusions

We demonstrated a novel way to produce trains of ultra-short electron bunches via interaction of two crossing laser pulses and their wakefields in plasma. For optimal experimental conditions, more than 50% of shots resulted in electron beams with up to 4 quasi-monoenergetic spectral peaks. Through the analysis of beam loading, we showed that these beams possess train structures, with individual bunches separated by a plasma wavelength. 2D PIC-simulations support the interpretation of our experimental results.

## Acknowledgments

We are grateful to Dr. Serge Kalmykov, Dr. Min Chen, and Dr. Miroslav Krůs for fruitful discussions and thank Kevin Brown and Chad Petersen for their help with the experiments and operating the laser system. We acknowledge Holland Computing Center at the University of Nebraska for providing computing infrastructure for the simulations. W.Y. acknowledges support by the High Field Initiative Project No. CZ.02.1.01/0.0/0.0/15_003/0000449. This material is based upon work supported by the National Science Foundation under Grant No. PHY-1535700 (ultra-low emittance electron beams), the US Department of Energy, High-Energy Physics, under Award # DE-SC0019421 (Controlled Injection of electrons for the Improved Performance of Laser-Wakefield Acceleration), and the US Department of Energy, Office of Science, Basic Energy Sciences, under Award # DE-FG02-05ER15663 (Laser-Driven X-rays for Ultrafast Science). This support does not constitute an express or implied endorsement on the part of the Government.

## References

[1] O. Lundh, J. Lim, C. Rechatin, L. Ammoura, a. Ben-Ismaïl, X. Davoine, G. Gallot, J.-P. Goddet, E. Lefebvre, V. Malka, and J. Faure, Nat. Phys. **7**, 219 (2011).

[2] G. Golovin, S. Banerjee, C. Liu, S. Chen, J. Zhang, B. Zhao, P. Zhang, M. Veale, M. Wilson, P. Seller, and D. Umstadter, Sci. Rep. **6**, 24622 (2016).

[3] G. Golovin, S. Chen, N. Powers, C. Liu, S. Banerjee, J. Zhang, M. Zeng, Z. Sheng, and D. Umstadter, Phys. Rev. Spec. Top. - Accel. Beams **18**, 011301 (2015).

[4] J. Wenz, A. Döpp, K. Khrennikov, S. Schindler, M.F. Gilljohann, H. Ding, J. Götzfried, A. Buck, J. Xu, M. Heigoldt, W. Helml, L. Veisz, and S. Karsch, Nat. Photonics **13**, 263 (2019).

[5] M. Hansson, B. Aurand, X. Davoine, H. Ekerfelt, K. Svensson, A. Persson, C.-G. Wahlström, and O. Lundh, Phys. Rev. Spec. Top. - Accel. Beams **18**, 071303 (2015).

[6] C. Thaury, E. Guillaume, A. Lifschitz, K. Ta Phuoc, M. Hansson, G. Grittani, J. Gautier, J.-P. Goddet, A. Tafzi, O. Lundh, and V. Malka, Sci. Rep. **5**, 16310 (2015).

[7] C. Aniculaesei, V.B. Pathak, K.H. Oh, P.K. Singh, B.R. Lee, C.I. Hojbota, T.G. Pak, E. Brunetti, B.J. Yoo, J.H. Sung, S.K. Lee, H.T. Kim, and C.H. Nam, Phys. Rev. Appl. **12**, 044041 (2019).

[8] Y. Wu, C. Yu, Z. Qin, W. Wang, R. Qi, Z. Zhang, K. Feng, L. Ke, Y. Chen, C. Wang, J. Liu, R. Li, and Z. Xu, Plasma Phys. Control. Fusion **61**, 085030 (2019).

[9] E. Allaria, F. Bencivenga, R. Borghes, F. Capotondi, D. Castronovo, P. Charalambous, P. Cinquegrana, M.B. Danailov, G. De Ninno, A. Demidovich, S. Di Mitri, B. Diviacco, D. Fausti, W.M. Fawley, E. Ferrari, L. Froehlich, D. Gauthier, A. Gessini, L. Giannessi, R. Ivanov, M. Kiskinova, G. Kurdi, B. Mahieu, N. Mahne, I. Nikolov, C. Masciovecchio, E. Pedersoli, G. Penco, L. Raimondi, C. Serpico, P. Sigalotti, S. Spampinati, C. Spezzani, C. Svetina, M. Trovò, and M. Zangrando, Nat. Commun. **4**, 2476 (2013).

[10] A. Gover, Phys. Rev. Spec. Top. - Accel. Beams **8**, 030701 (2005).

[11] Y.-C. Huang, Int. J. Mod. Phys. B **21**, 287 (2007).

[12] C. Jing, A. Kanareykin, J.G. Power, M. Conde, Z. Yusof, P. Schoessow, and W. Gai, Phys. Rev. Lett. **98**, 144801 (2007).

[13] P. Muggli, V. Yakimenko, M. Babzien, E. Kallos, and K.P. Kusche, Phys. Rev. Lett. **101**, 054801 (2008).

[14] Y.-E. Sun, P. Piot, A. Johnson, A.H. Lumpkin, T.J. Maxwell, J. Ruan, and R. Thurman-Keup, Phys. Rev. Lett. **105**, 234801 (2010).

[15] C.M.S. Sears, E. Colby, R. Ischebeck, C. McGuinness, J. Nelson, R. Noble, R.H. Siemann, J. Spencer, D. Walz, T. Plettner, and R.L. Byer, Phys. Rev. Spec. Top. - Accel. Beams **11**, 061301 (2008).

[16] M. Boscolo, M. Ferrario, I. Boscolo, F. Castelli, and S. Cialdi, Nucl. Instruments Methods Phys. Res. Sect. A Accel. Spectrometers, Detect. Assoc. Equip. **577**, 409 (2007).

[17] A. Oguchi, A. Zhidkov, K. Takano, E. Hotta, K. Nemoto, and K. Nakajima, Phys. Plasmas **15**, 043102 (2008).

[18] M. Zeng, M. Chen, L.L. Yu, W.B. Mori, Z.M. Sheng, B. Hidding, D.A. Jaroszynski, and J. Zhang, Phys. Rev. Lett. **114**, 084801 (2015).

[19] S.Y. Kalmykov, X. Davoine, I. Ghebregziabher, and B.A. Shadwick, Nucl. Instruments Methods Phys. Res. Sect. A Accel. Spectrometers, Detect. Assoc. Equip. **909**, 433 (2018).

[20] S.Y. Kalmykov, X. Davoine, I. Ghebregziabher, R. Lehe, A.F. Lifschitz, and B.A. Shadwick, Plasma Phys. Control. Fusion **58**, 034006 (2016).

[21] S.Y. Kalmykov, I.A. Ghebregziabher, X. Davoine, R. Lehe, A.F. Lifschitz, V. Malka, and B.A. Shadwick, AIP Conf. Proc. **1777**, 080007 (2016).

[22] Z. Lécz, A. Andreev, I. Konoplev, A. Seryi, and J. Smith, Plasma Phys. Control. Fusion **60**, 075012 (2018).

[23] O. Lundh, C. Rechatin, J. Lim, V. Malka, and J. Faure, Phys. Rev. Lett. **110**, 065005 (2013).

[24] M.R. Islam, E. Brunetti, R.P. Shanks, B. Ersfeld, R.C. Issac, S. Cipiccia, M.P. Anania, G.H. Welsh, S.M. Wiggins, A. Noble, R.A. Cairns, G. Raj, and D.A. D A Jaroszynski, New J. Phys. **17**, 093033 (2015).

[25] M. Heigoldt, A. Popp, K. Khrennikov, J. Wenz, S.W. Chou, S. Karsch, S.I. Bajlekov, S.M. Hooker, and B. Schmidt, Phys. Rev. Spec. Top. - Accel. Beams **18**, 121302 (2015).

[26] G. Golovin, W. Yan, J. Luo, C. Fruhling, D. Haden, B. Zhao, C. Liu, M. Chen, S. Chen, P. Zhang, S. Banerjee, and D. Umstadter, Phys. Rev. Lett. **121**, 104801 (2018).

[27] G. Golovin, S. Banerjee, S. Chen, N. Powers, C. Liu, W. Yan, J. Zhang, P. Zhang, B. Zhao, and D. Umstadter, Nucl. Instruments Methods Phys. Res. Sect. A Accel. Spectrometers, Detect. Assoc. Equip. **830**, 375 (2016).

[28] T.C. Katsouleas, J.J. Su, S. Wilks, J.M. Dawson, and P. Chen, Part. Accel. **22**, 81 (1987).

[29] M. Tzoufras, W. Lu, F. Tsung, C. Huang, W. Mori, T. Katsouleas, J. Vieira, R. Fonseca, and L. Silva, Phys. Rev. Lett. **101**, 145002 (2008).

[30] T.D. Arber, K. Bennett, C.S. Brady, A. Lawrence-Douglas, M.G. Ramsay, N.J. Sircombe, P. Gillies, R.G. Evans, H. Schmitz, A.R. Bell, and C.P. Ridgers, Plasma Phys. Control. Fusion **57**, 113001 (2015).

[31] F.S. Tsung, W. Lu, M. Tzoufras, W.B. Mori, C. Joshi, J.M. Vieira, L.O. Silva, and R.A. Fonseca, Phys. Plasmas **13**, 056708 (2006).

[32] A.A. Golovanov and I.Y. Kostyukov, Phys. Plasmas **25**, 103107 (2018).

[33] V. Horný, V. Petržílka, O. Klimo, and M. Krůs, Phys. Plasmas **24**, 103125 (2017).

[34] H. Kotaki, S. Masuda, M. Kando, J.K. Koga, and K. Nakajima, Phys. Plasmas **11**, 3296 (2004).

[35] G. Fubiani, E. Esarey, C.B. Schroeder, and W.P. Leemans, Phys. Rev. E **70**, 016402 (2004).

[36] Z.M. Sheng, W.M. Wang, R. Trines, P. Norreys, M. Chen, and J. Zhang, Eur. Phys. J. Spec. Top. **175**, 49 (2009).

[37] D. Umstadter, J. Kim, and E. Dodd, Phys. Rev. Lett. **76**, 2073 (1996).